\RequirePackage{fix-cm}
\documentclass[smallcondensed]{svjour3}     % onecolumn (ditto)
\smartqed  % flush right qed marks, e.g. at end of proof
\usepackage{graphicx}
\usepackage{mathptmx}      % use Times fonts if available on your TeX system
\begin{document}

\title{A three dimensional investigation of two dimensional orbits}
%\subtitle{Do you have a subtitle?\\ If so, write it here}

%\titlerunning{Short form of title}        % if too long for running head

\author{D. D. Carpintero \and J. C. Muzzio}

%\authorrunning{Short form of author list} % if too long for running head

\institute{D. D. Carpintero \at
Fac. de Ciencias Astron\'omicas y Geof\'\i sicas,
Universidad Nacional de La Plata and Instituto de Astrof\'\i sica
de La Plata, CCT Conicet La Plata - UNLP \\
Tel.: +54-221-423-6593 \\
\email{ddc@fcaglp.unlp.edu.ar} 
           \and
J. C. Muzzio \at
Fac. de Ciencias Astron\'omicas y Geof\'\i sicas,
Universidad Nacional de La Plata and Instituto de Astrof\'\i sica
de La Plata, CCT Conicet La Plata - UNLP \\
Tel.: +54-221-423-6593 \\
\email{jcmuzzio@fcaglp.unlp.edu.ar} 
}

\date{Received: date / Accepted: date}
% The correct dates will be entered by the editor

\maketitle

\begin{abstract}

Orbits in the principal planes of triaxial potentials are known to be prone to
unstable motion normal to those planes, so that three dimensional investigations
of those orbits are needed even though they are two dimensional. We present here
an investigation of such orbits in the well known logarithmic potential  which
shows that the third dimension must be taken into account when studying them and
that the instability worsens for lower values of the forces normal to the plane.
Partially chaotic orbits are present around resonances, but also in other
regions. The action normal to the plane seems to be related to the isolating
integral that distinguishes regular form partially chaotic orbits, but not to
the integral that distinguishes partially from fully chaotic orbits.

\end{abstract}

\section{Introduction}
\label{sec:intro}

Stellar systems are clearly three-dimensional (3D) in configuration space.
Nevertheless, since the study of two-dimensional (2D) systems is much easier
than that of 3D ones, both analiytically and numerically, it is tempting in many
instances to resort to the 2D approach. Standard examples are investigations of
2D orbits in the equatorial plane of disk galaxies or in the principal planes of
elliptical ones. The risks of resorting to this simplification in the case of
elliptical galaxies were emphasized by \cite{MF96}, but their words of caution
may be valid for disk galaxies as well. The main problem is that orbits in the
principal planes are often unstable to perturbations out of the plane, so that
such instability would not be detected in 2D studies and, as a result, orbits
that a full 3D study would reveal as chaotic might appear as regular in a 2D
investigation.\footnote{One property of chaotic motion is the exponential
divergence of orbits (which also characterizes unstable motion) while, at the
same time, the motion is bound (i.e., an hyperbolic orbit in a Newtonian field
is unstable, but not chaotic). Here we deal exclusively with bound orbits, so
that we will use the terms \emph{chaotic} and \emph{unstable} indistinctly.}
Another investigation of orbits in triaxial models with weak cusps by
\cite{FM97} also found many cases of instability normal to the principal planes.
Besides, the more recent investigation of \cite{Aetal07} puts the original
warning of \cite{MF96} in an even more general context.

There is a mathematical peculiarity that makes a full 3D study of 2D orbits in
the principal planes particularly interesting. Let us assume that we choose the
$(x, y)$ plane and that we adopt initial conditions $x = x_0$, $y = y_0$, $z =
0$, $V_x = V_{x0}$, $V_y = V_{y0}$ and $V_{z} = 0$. Since both $z$ and $V_z$ are
exactly zero and there is no rounding off error in the numerical representation
of that number, the time derivatives of $z$ and $V_z$ in the equations of motion
will remain always equal to zero and the orbit will remain forever on the $(x,
y)$ plane, even if it is unstable in the $z$ direction. Nevertheless, if the
variational equations are computed at the same time as the orbit, they will
clearly reveal how an infinitesimally small departure from that orbit will
exponentially grow and, thus, the chaotic nature of the orbit. Therefore, a 3D
investigation of 2D orbits on the principal planes of a triaxial potential
computing the Lyapunov exponents via the variational equations seems warranted.

Moreover, we have emphasized several times over the past decade (e.g.
\cite{Muz03}, \cite{MM04}, \cite{MCW05}, \cite{AMNZ07}, \cite{MNZ09}) the
importance of distinguishing between partially and fully chaotic orbits in
dynamical studies of stellar systems. The sole isolating integral of fully
chaotic orbits is the energy, while partially chaotic orbits obey an additional
isolating integral or pseudo integral. The Lyapunov exponents offer a simple way
to distinguish these orbits because, due to the conservation of phase space
volume in Hamiltonian systems, these exponents come in pairs, one positive and
one negative, with the same absolute value. Besides, energy conservation in
autonomous systems ensures that one of those pairs has zero absolute value, and
each additional isolating integral results in another pair of exponents equal to
zero. Therefore, in a 3D potential we may have all three pairs of exponents
equal to zero (regular orbit), only one pair not equal to zero (partially
chaotic orbit), or two non--zero pairs of Lyapunov exponents (fully chaotic
orbit). In 2D potentials, instead, we can only have regular (two null pairs of
exponents) or partially chaotic orbits (one null and one non--null pairs).

There has been some discussion recently about whether the partially chaotic
orbits are simply those that belong in the stochastic layer surrounding
resonances around regular orbits (\cite{NPM07}) or they cover more extended
regions of phase space (\cite{AMNZ07}). Now, 2D orbits can be either regular or
partially chaotic, but not fully chaotic because they can obey two integrals of
motion only or, alternatively, they have only four Lyapunov exponents
instead of the six of 3D orbits. Thus, we can expect that an investigation of
orbits close to the main planes of a triaxial potential may shed some light on
the nature of the partially chaotic orbits.

The logarithmic potential (see, e.g. \cite{BT08}) has been frequently used for
2D studies (see, e.g. \cite{MS89}, \cite{PL96}, \cite{CA98}, \cite{FB01}), so
that it may be interesting to check how much it is affected by instabilities out
of the plane. Besides, it has a simple mathematical expression that allows a
reasonably fast computation of all the Lyapunov exponents. Therefore, we decided
to perfom an investigation of 2D orbits on the principal planes of the 3D
logarithmic potential. The next section describes our method and the third one
presents our results which are then discussed in the fourth section.

\section{Method}
\label{sec:method}

Our investigation is based on the comparison of the chaoticity of orbits moving
in a 2D potential with that of orbits moving in a 3D potential. Two
possibilities are considered in the latter case: 1) Initial conditions very
close to one of the principal planes of symmetry, but not exactly on it; 2)
Initial conditions precisely on a principal plane. The reason for considering
these two  possibilities is that, as indicated above, in the second case the
orbit is forced to remain on the plane for mathematical reasons, even if it is
unstable perpendicularly to the plane. In order to recognize regular from
chaotic, and partially from fully chaotic orbits, we compute all the Lyapunov
exponents corresponding to the case in question, i.e., four in the 2D, and six
in the 3D cases.

As indicated, we adopted the 3D logarithmic potential:
\begin{equation}
\Phi(\vec{x})={V_0^2\over 2}\ln\left({R_{\rm c}^2}+{x^2\over a^2}+{y^2\over
b^2}+{z^2\over c^2}\right), 
\end{equation}
where $V_0$, $R_{\rm c}$, $a$, $b$ and $c$ are constants. The equipotentials are
ellipsoids whose semiaxes are proportional to $a$, $b$ and $c$. For the  2D case
we simply omitted the $z$ coordinate, both in the equations of motion and in the
variational equations.

In order to compare corresponding orbits in the 2D and 3D cases, we chose a
fixed set of initial conditions. We took 90 equidistant points along each axis
of the $\{y,V_y\}$ plane defined by $E=0$ and $x=z=V_z=0$, where $E$ is the
energy of the orbit. The allowed region on this plane is defined by the
condition $V_x^2>0$. The symmetries of the potential allowed us to restrict our
study to the region $y>0$, $V_x>0$ and $V_y>0$. We then set $V_0=1$, $R_{\rm
c}=0.01$, $a=1$ and $b=0.7$ and we picked up each of the resulting $(y,V_y)$
pairs as initial conditions whenever they allowed to obtain $E=0$ with a real
value of $V_x$.

We integrated the equations of motion and the corresponding variational
equations with a double precision Runge-Kutta-Fehlberg algorithm of order 7,
with Cash-Karp coefficients \cite{PTVF86}. In order to obtain information about
the degree of chaoticity of the orbits, the full set of Lyapunov exponents was
computed for each orbit using the well-known algorithm developed by
\cite{BGGS80a}, \cite{BGGS80b}. To this end, the six necessary variational
initial conditions were obtained by perturbing the original ones along each of
the six axes of the phase space. The orbits were integrated over $t=10,000$ time
units (t.u.), which corresponds to a number of orbital periods between 1500 and
3500, approximately. The energy is conserved to better than $3\times 10^{-10}$
in all cases, and better than $5\times 10^{-11}$ in most cases.

\begin{figure}
 \includegraphics[width=0.5\textwidth]{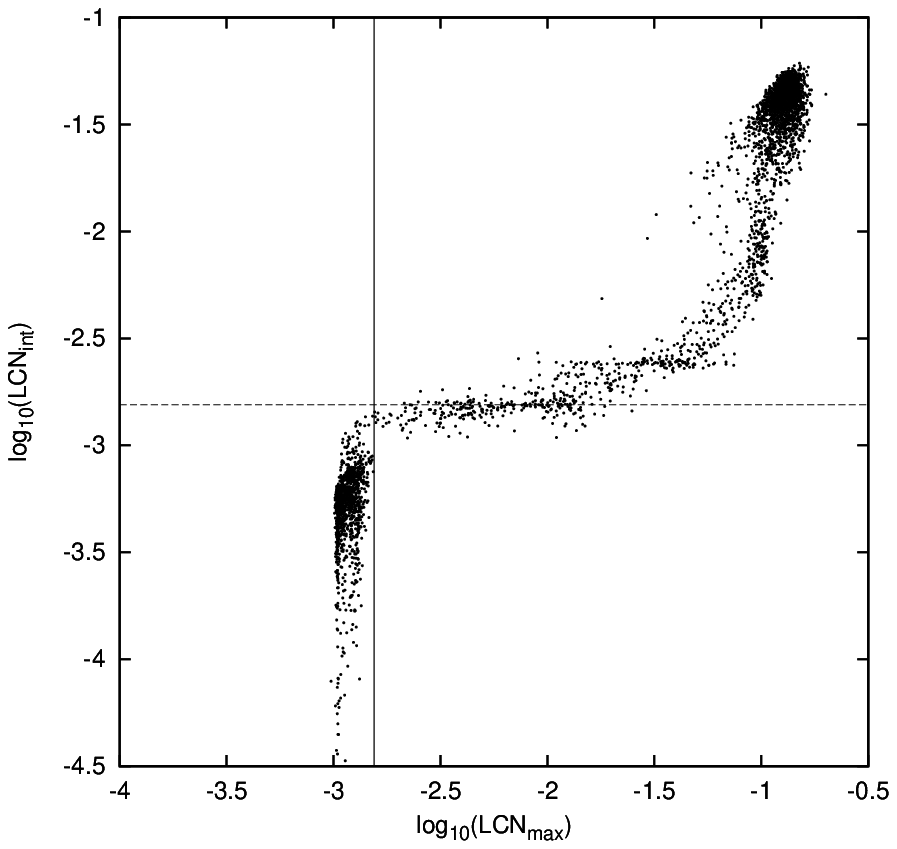}
 \includegraphics[width=0.5\textwidth]{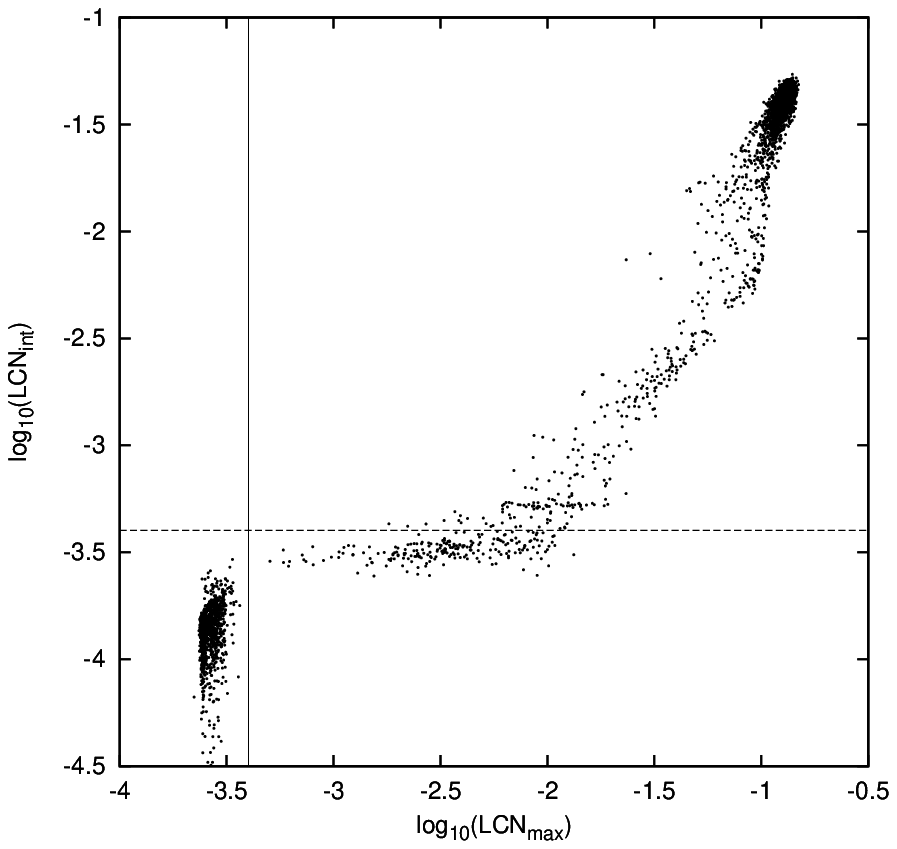}
 \caption{Largest Lyapunov characteristic numbers (LCN$_{\rm max}$) vs.
intermediate Lyapunov characteristic numbers (LCN$_{\rm int}$) for the $c=0.50$
potential studied. The integration time was 10,000 t.u. for the left plot and
50,000 t.u. for the right one. The limiting value ${\rm LCN}=0.00155$ is marked
with straight lines as a reference on the left plot; the corresponding limiting
value ${\rm LCN}=0.00040$ is marked on the right one.}
 \label{fig:lcns05}
\end{figure}

As is well known, the Lyapunov exponents, being defined as values of a certain
function $f(t)$ when $t\to\infty$, cannot be computed directly from a numerical
experiment. Instead, the Lyapunov characteristic numbers (LCNs) are used, that
is, the values of $f(t)$ at finite times, the validity of this approximation
resting on the reasonable assumption that the LCNs will tend to the Lyapunov
exponents as $t\to\infty$. However, this raises the question of what LCN value
will be considered the equivalent of a zero value of the corresponding Lyapunov
exponent. To answer it we present in Fig. \ref{fig:lcns05} (left) a plot of the
largest LCN, LCN$_{\rm max}$, versus the second largest one, LCN$_{\rm int}$ for
the case with $c = 0.5$ (see below). The figure clearly shows a cluster of
points near the left lower corner that corresponds to regular orbits and another
cluster near the right upper corner that corresponds to the fully chaotic
orbits. Besides, there is an horizontal band that extends over most of the
middle part of the plot that corresponds to the partially chaotic orbits. The
horizontal and vertical lines at ${\rm LCN}=0.00155$ show the chosen limiting
value for the LCNs. No doubt, that limiting value is only approximate and of
statistical nature, since a few orbits with LCNs slightly larger (smaller) than
the limiting value might actually be regular (chaotic), but the figure shows
that it clearly separates the three different types of orbits. The right part of
Fig. \ref{fig:lcns05} shows the same plot as the left part, but using an
integration time of 50,000 t.u. The limiting value is lower now (about ${\rm
LCN}=0.00040$) but, except for that fact, the general aspect of both plots is
essentially the same. Of course, some sticky orbits that appear as regular on
the left figure reveal their true chaotic nature on the right one, but the
effect is small (the number of regular orbits falls from $29.2\%$ to $28.7\%$
only), so that the chosen integration limit of 10,000 t.u. is most adequate for
the present investigation.  

With these tools, we integrated orbits in several potentials with the
abovementioned fixed values of $V_0$, $R_{\rm c}$, $a$ and $b$. In the 2D case
nothing more is needed. In the 3D case, however, the parameter $c$ is still
undefined. We took several values of $c$ in the interval $(0,1]$. Thus, the
series of studied potentials goes from very flattened and similar to the 2D
case, through axisymmetric prolate (when $c=0.7$), to axisymmetric oblate (when
$c=1$). This allowed us to study whether and when the third dimension, although
not seen by the star, plays a role in its regularity or chaoticity.

\section{Numerical experiments and results}
\label{sec:results}

\begin{figure}
 \includegraphics[width=0.5\textwidth]{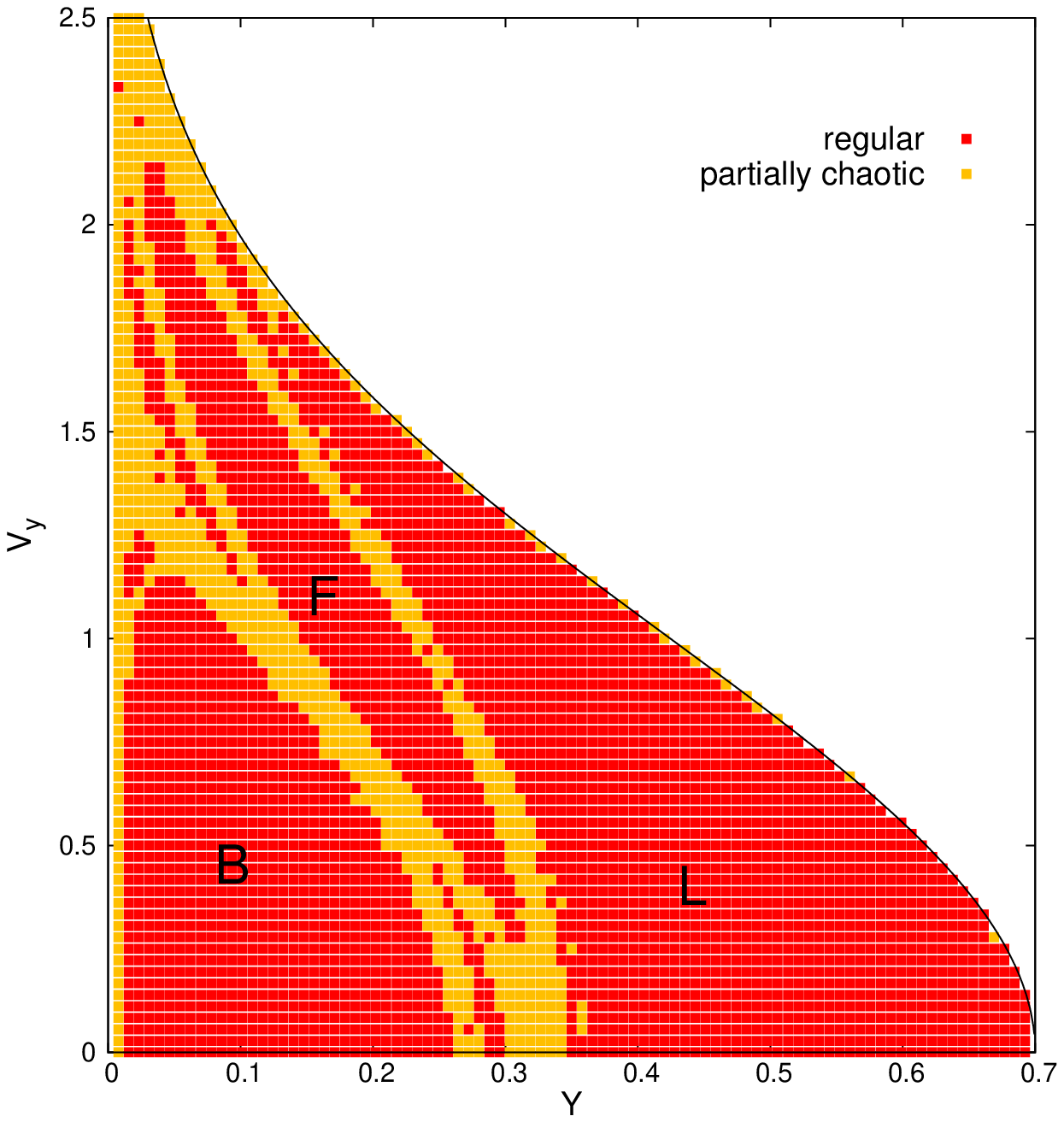}
 \includegraphics[width=0.5\textwidth]{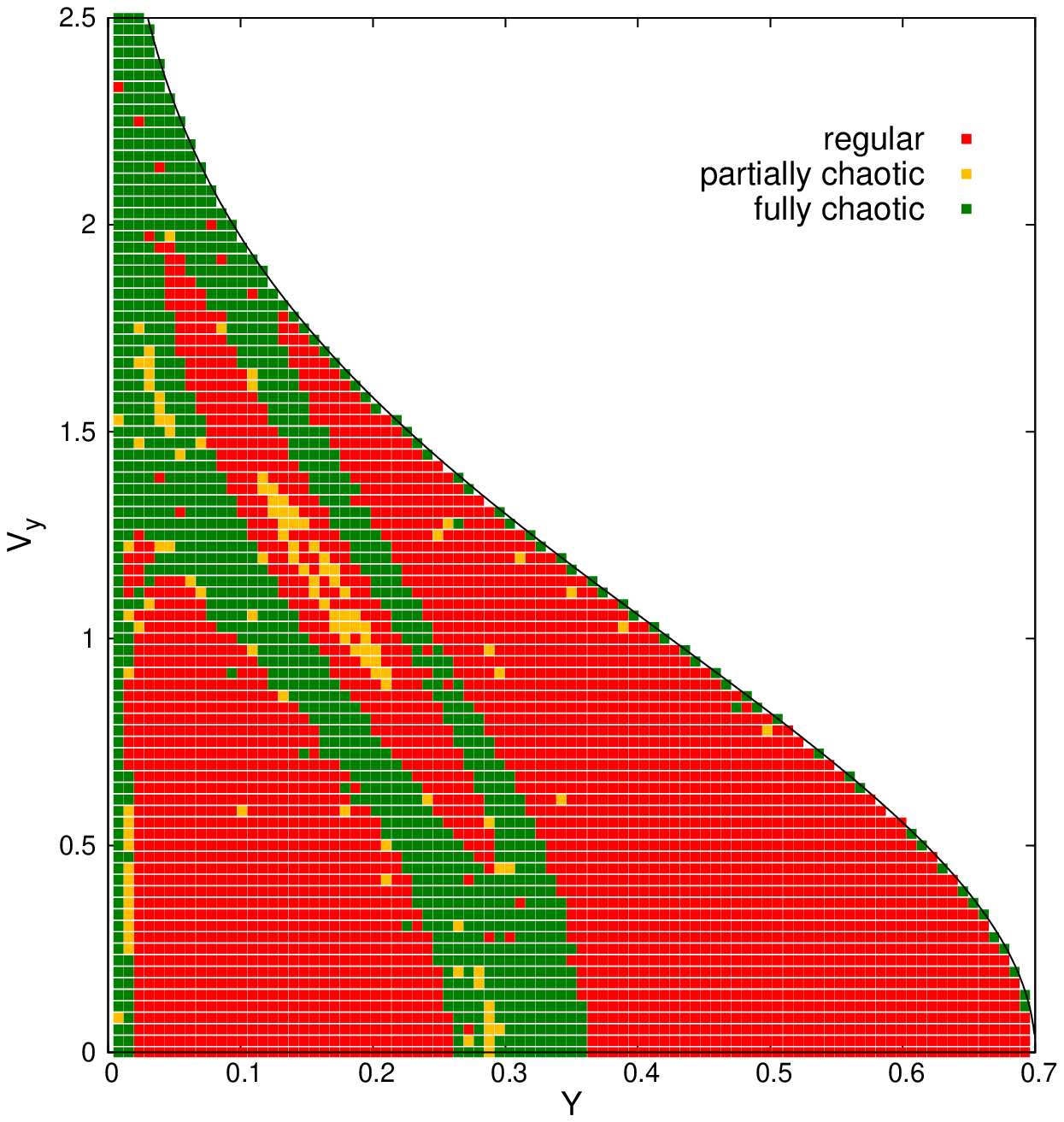}
 \caption{Left: Regular or partially chaotic character of orbits in the 2D
potential model, each of which is launched with initial conditions $E=x=0$ and
$(y,V_y)$ given by the coordinates of the corresponding point on the plot.
White regions correspond to regular orbits;  letters into the plot refer to the
main regular families (see text) and the limiting ($V_x=0$) curve is also
shown. Right: Regular, partially or fully chaotic character of orbits with the
same initial conditions of the left figure, plus $z=10^{-6}, V_z=0$, in the
3D potential model with $c=0.30$.}
 \label{fig:2D-3D03}
\end{figure}

We first reproduced (Fig. \ref{fig:2D-3D03}, left) the well-known orbital
content of the 2D potential model \cite{MS89}.

As indicated before, the only chaotic orbits here are the partially chaotic ones
and the connected region of those orbits, i.e., those having the energy as the
only isolating integral of motion, is clearly seen. The regular orbits are
separated into disjoint regions, each one hosting a different family spawned by
a specific mother closed orbit. Although we do not need to specify the families
for the purposes of the present paper, we will mention the main ones here to be
able to identify the regions of the plot later on. The region marked "L" in Fig.
\ref{fig:2D-3D03} (left) is occupied by loop orbits, i.e., regular orbits in
which the frequencies of greatest amplitude along each axis are in a 1:1 ratio.
The "B" region in the figure corresponds to banana orbits, with a 2:1 ratio
between the abovementioned frequencies. Finally, the region with an "F" is
filled with fish orbits, with a 3:2 ratio. (It is worth noticing that these
rational ratios {\em are not} ratios between the fundamental frequencies of the
corresponding orbital tori.)

\begin{figure}
 \includegraphics[width=0.5\textwidth]{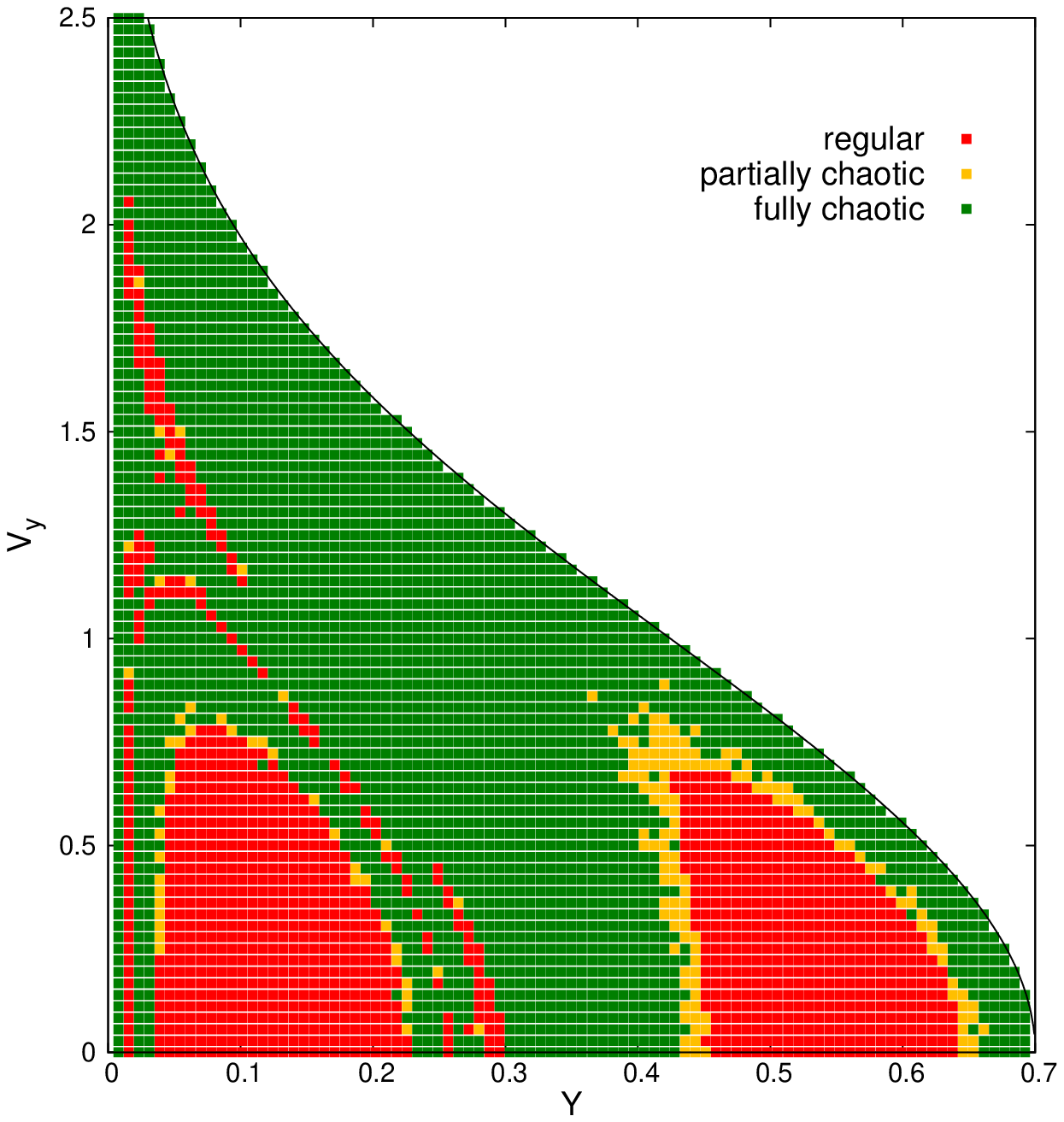}
 \includegraphics[width=0.5\textwidth]{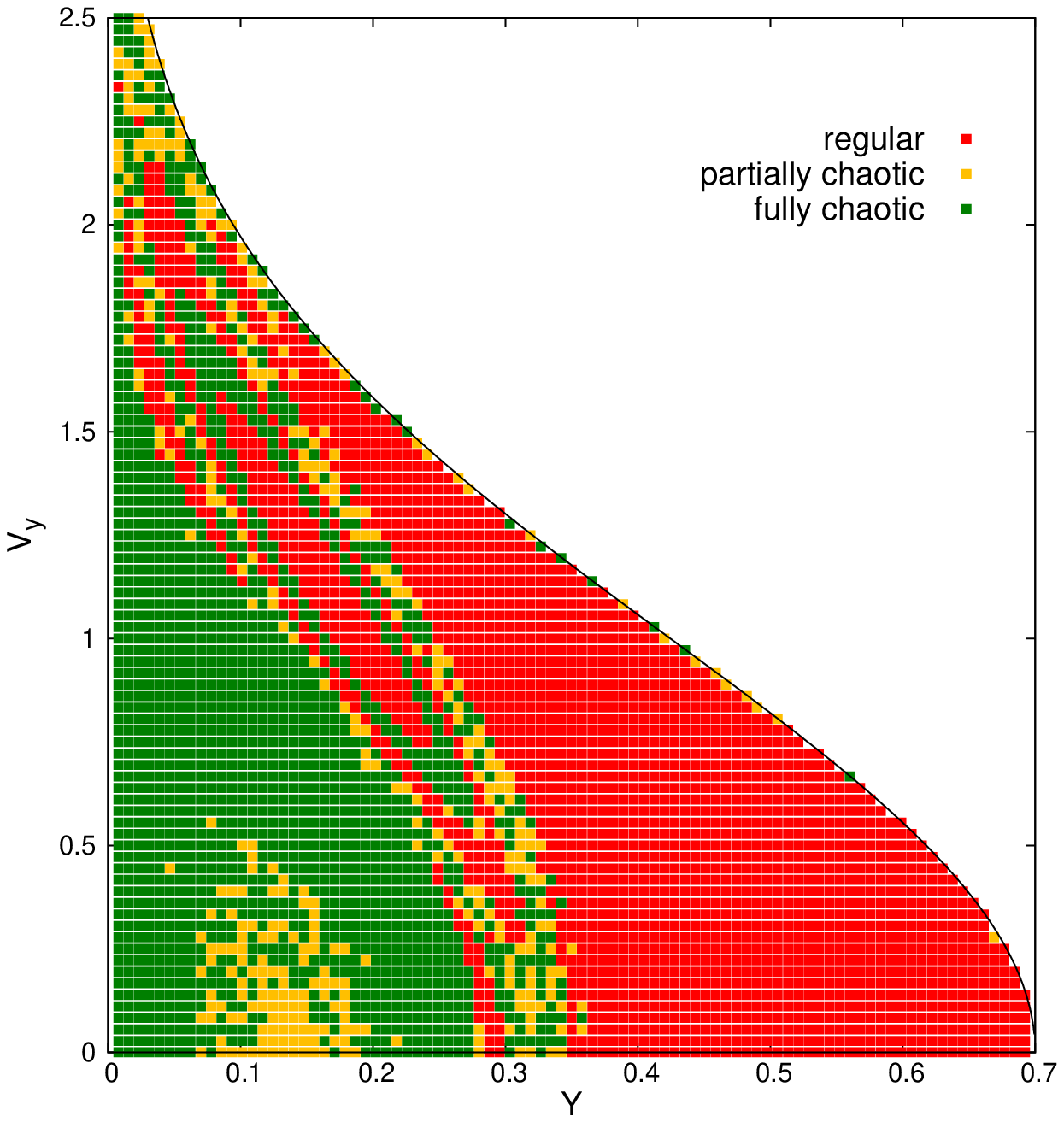}
 \caption{Left: Regular, partially or fully chaotic character of orbits in the
3D potential model with $c=0.50$. Right: Regular, partially or fully chaotic
character of orbits in the 3D potential model with $c=0.695$.}
 \label{fig:3D05-3D0695}
\end{figure}

\begin{figure}
 \includegraphics[width=0.5\textwidth]{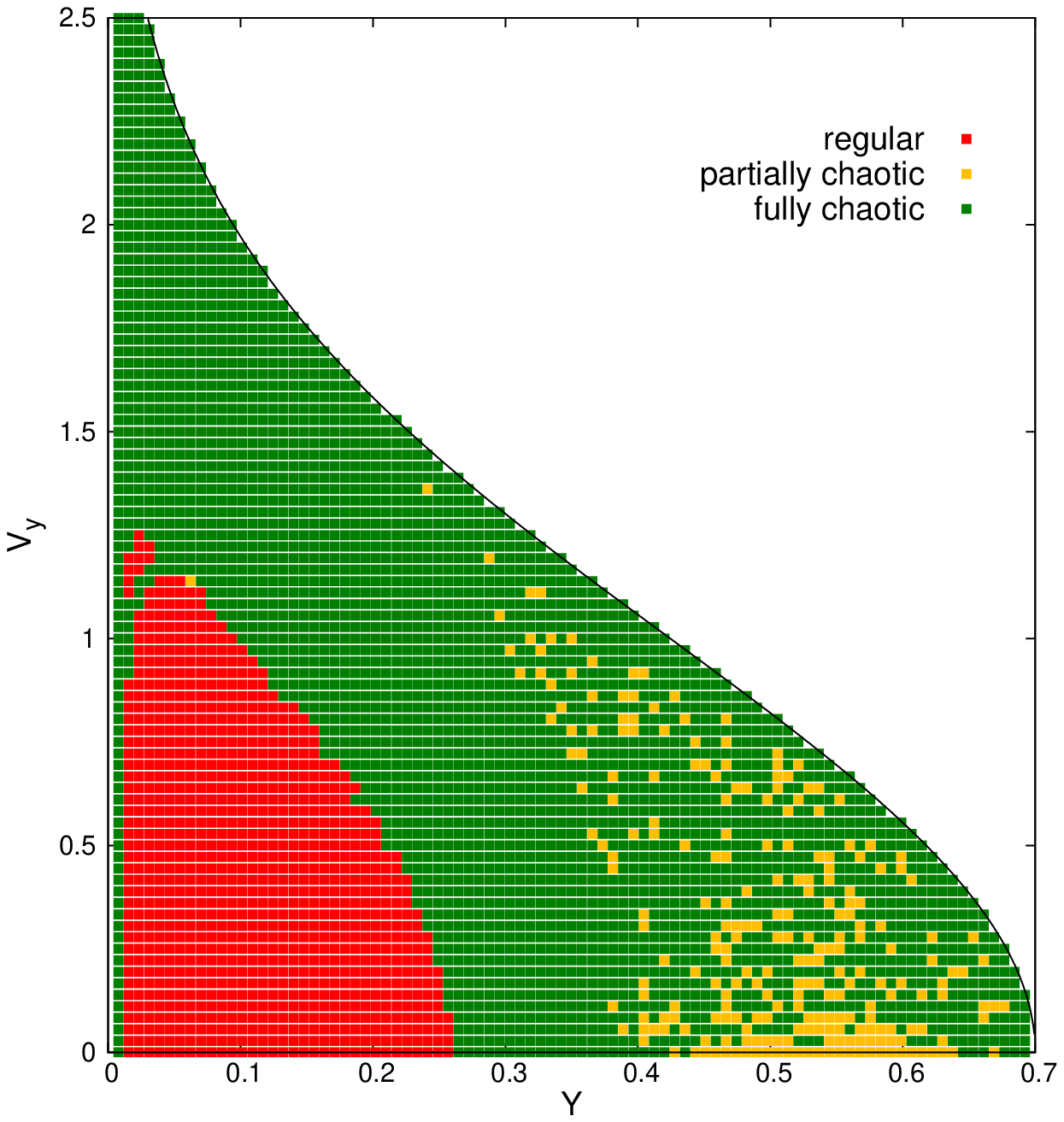}
 \includegraphics[width=0.5\textwidth]{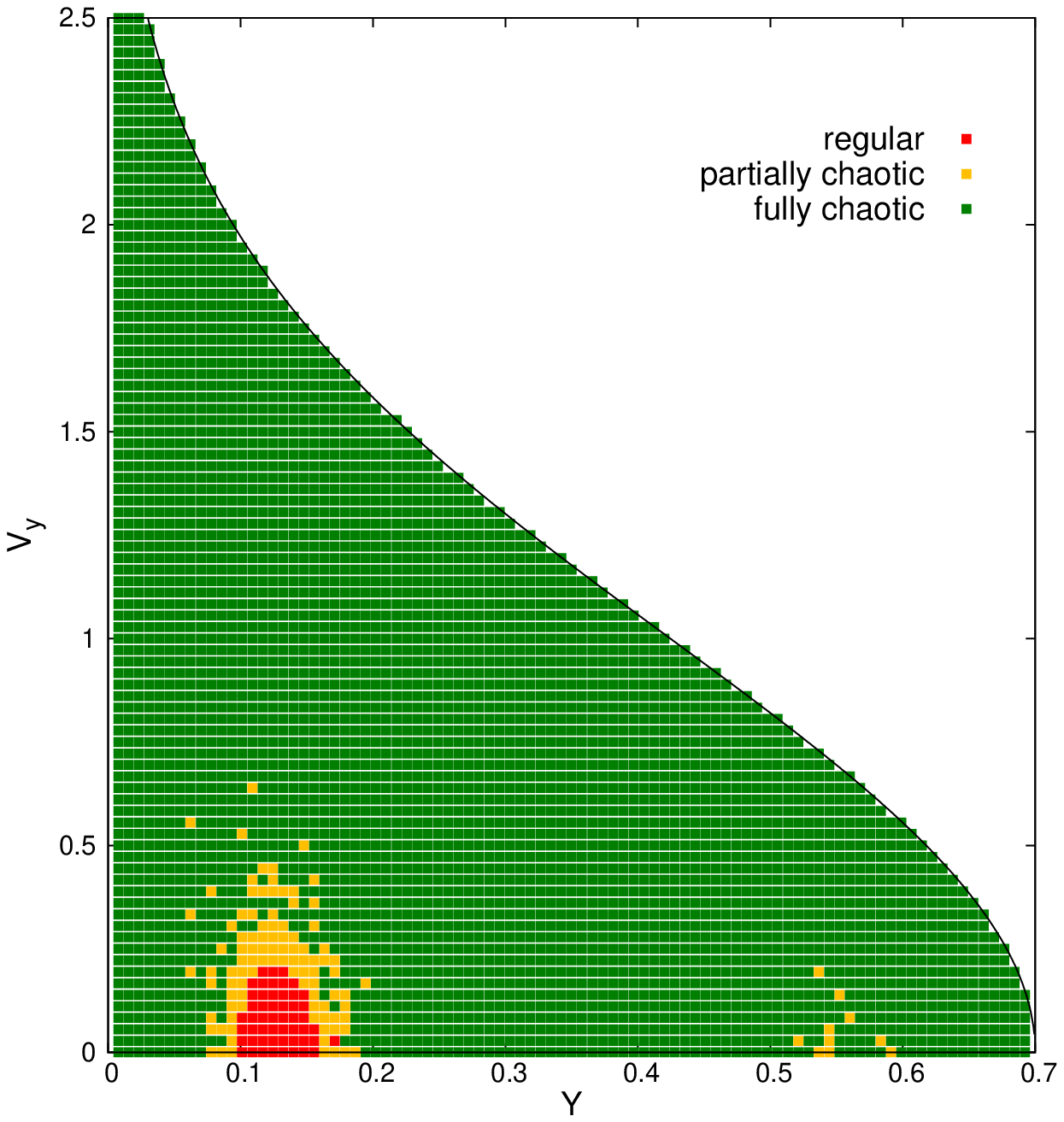}
 \caption{Left: Regular, semi-chaotic or chaotic character of orbits in the 3D
potential model with $c=0.705$. Right: the same, for the case $c=0.800$.}
 \label{fig:3D0705-3D08}
\end{figure}

Next we integrated orbits in the 3D potential for different $c$ values. We used
the same initial conditions as for the 2D potential, plus $z=10^{-6}, V_z=0$.
For small $c$ values, there is no much difference with the 2D results shown in
Fig. \ref{fig:2D-3D03} (left), e.g., barely 4 points switch from partially to
fully chaotic for $c=0.05$, and that number rises only to 42 points for
$c=0.20$. As the value of $c$ is increased further, however, more partially
chaotic orbits switch to fully chaotic and a few regular orbits switch to
partially chaotic. The right panel of Fig. \ref{fig:2D-3D03} presents the
results for $c=0.30$, when all the partially chaotic orbits have turned to fully
chaotic and a small island of partially chaotic orbits has appeared inside the
regular region corresponding to the fish orbits. As we continued increasing $c$
the region covered by fully chaotic orbits extends its area and other regions of
partially chaotic orbits appear as shown by Fig. \ref{fig:3D05-3D0695} (left)
that presents the results for $c=0.50$. Nevertheless, it is obvious that the
situation should change when approaching $c=b=0.70$, because then we have a
prolate potential that conserves the component of the angular momentum parallel
to the major axis and, thus, no fully chaotic orbits can be present. In fact,
the corresponding figure for that case is essentially the same as the one
presented in Fig. \ref{fig:2D-3D03} (left) for the 2D case. But, very close to
$c=0.70$, the situation is different, as shown in Fig. \ref{fig:3D05-3D0695}
(right), corresponding to $c=0.695$ and Fig. \ref{fig:3D0705-3D08} (left), for
the case $c=0.705$. In both cases, most of the initial conditions result in
chaotic orbits, mainly fully chaotic ones, but the regular orbits that originate
from initial conditions on the $(y, V_y)$ plane, when $y$ is still the smallest
axis ($c=0.695$) tend to be loops, while when $y$ becomes the intermediate axis
those regular orbits are mainly bananas. This result is quite natural, because
there are no stable tubes around the intermediate axis.

As $c$ is increased even further the bananas last "island of resistance" shrinks
and the few scattered partially chaotic orbits in the (former) tubes region
transform in fully chaotic orbits, as shown by Fig. \ref{fig:3D0705-3D08}
(right). As we approach the value $c=1.00$, chaos recedes because for that value
we get again a rotationally symmetric (prolate) potential, and we get again a
plot similar to that of Fig. \ref{fig:2D-3D03} (left).

\begin{figure}
 \includegraphics[width=0.5\textwidth]{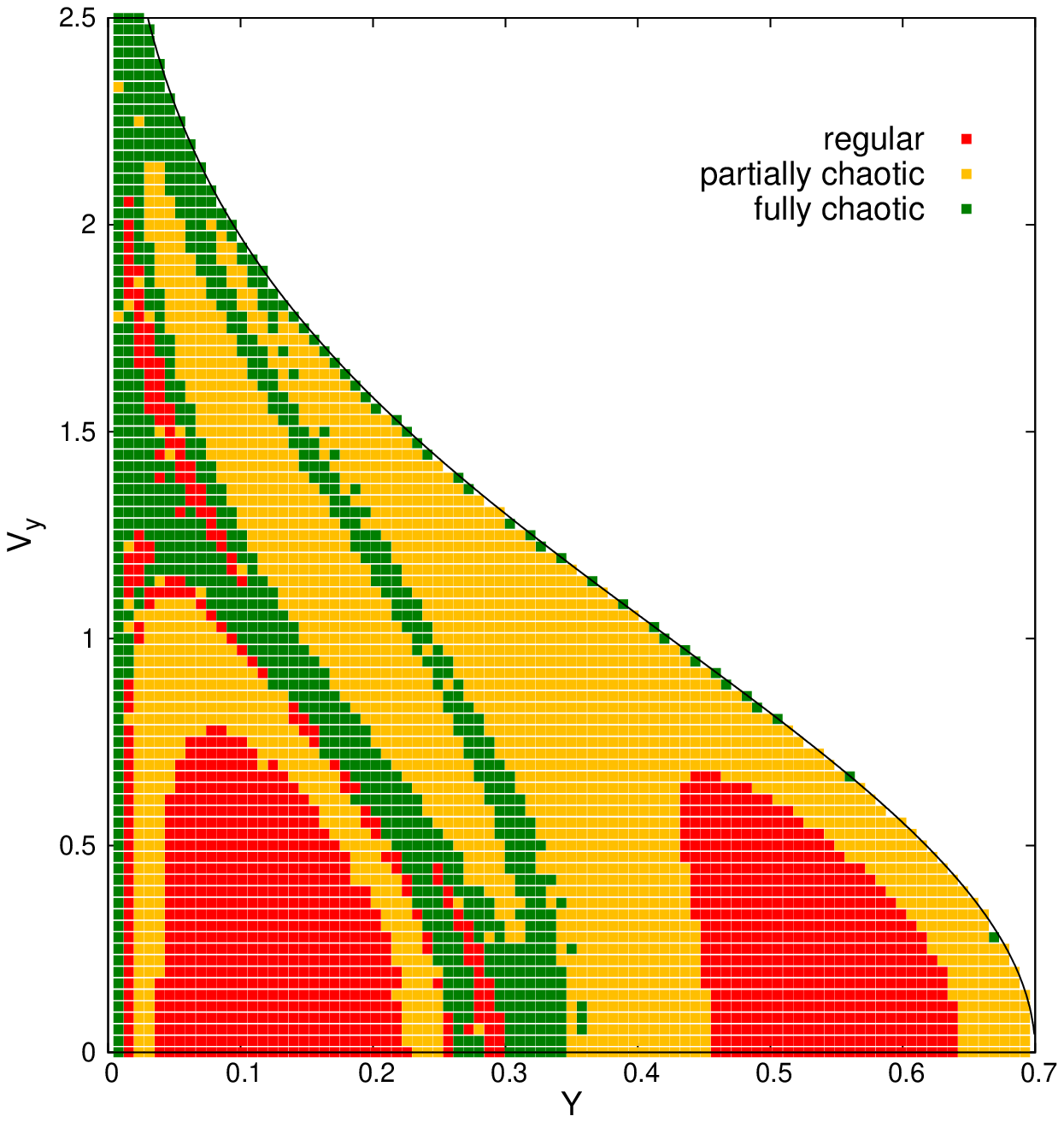}
 \includegraphics[width=0.5\textwidth]{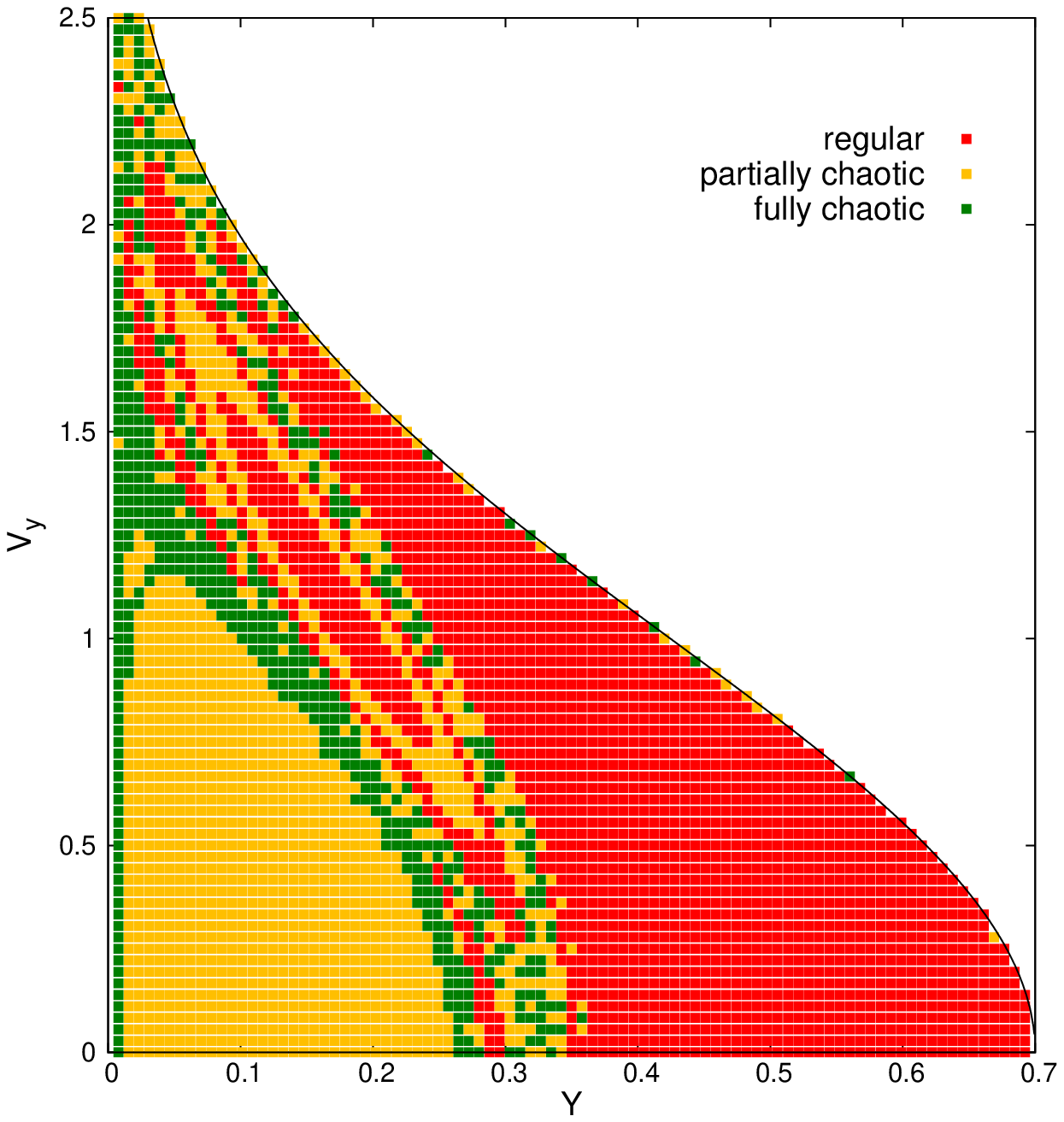}
 \caption{Left: Regular, semi-chaotic or chaotic character of orbits in the 3D
potential model with $c=0.500$ when $z=0$. Right: the same, for the case
$c=0.695$.}
 \label{fig:3D05-0695z0}
\end{figure}

\begin{figure}
 \includegraphics[width=0.5\textwidth]{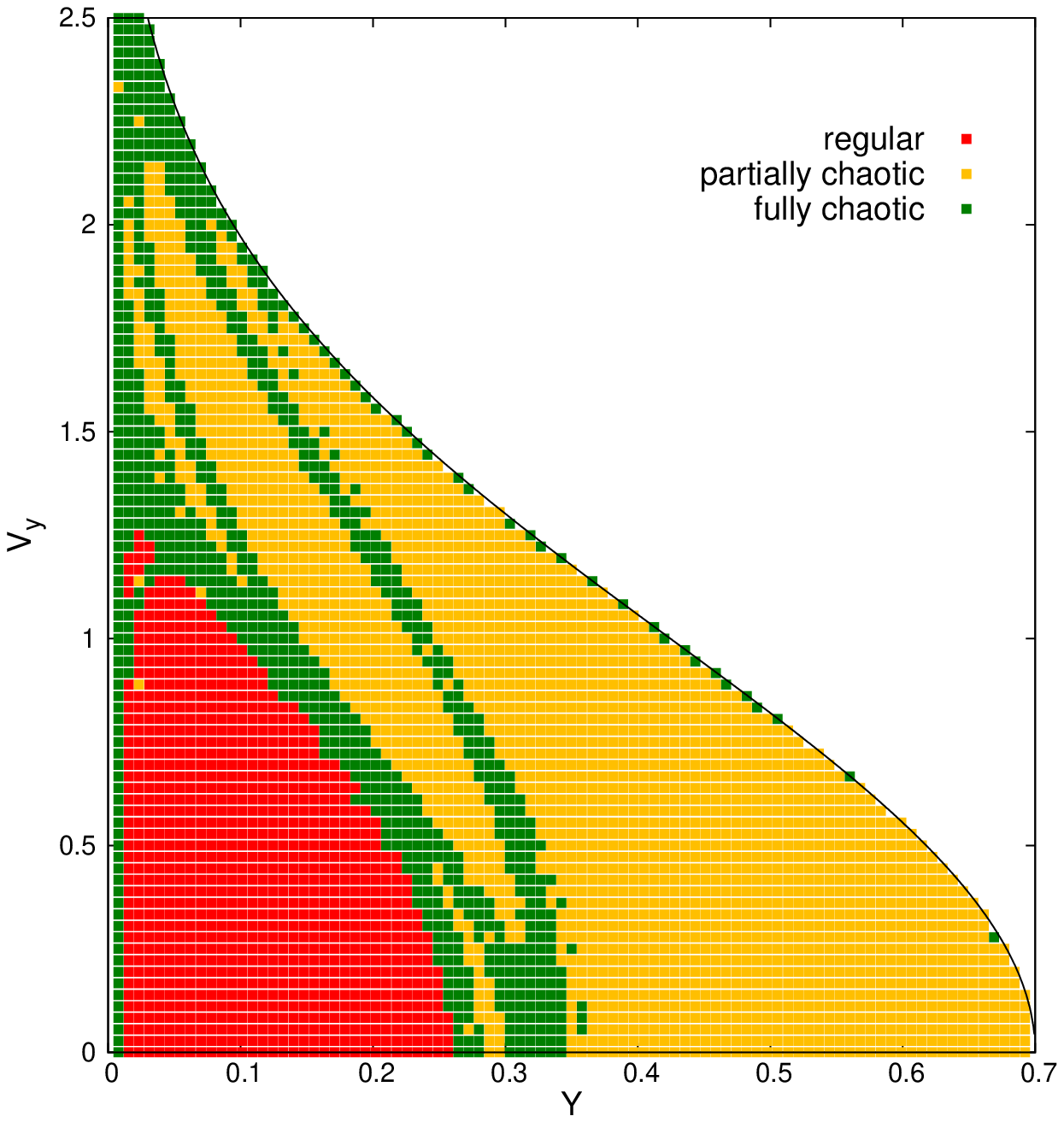}
 \includegraphics[width=0.5\textwidth]{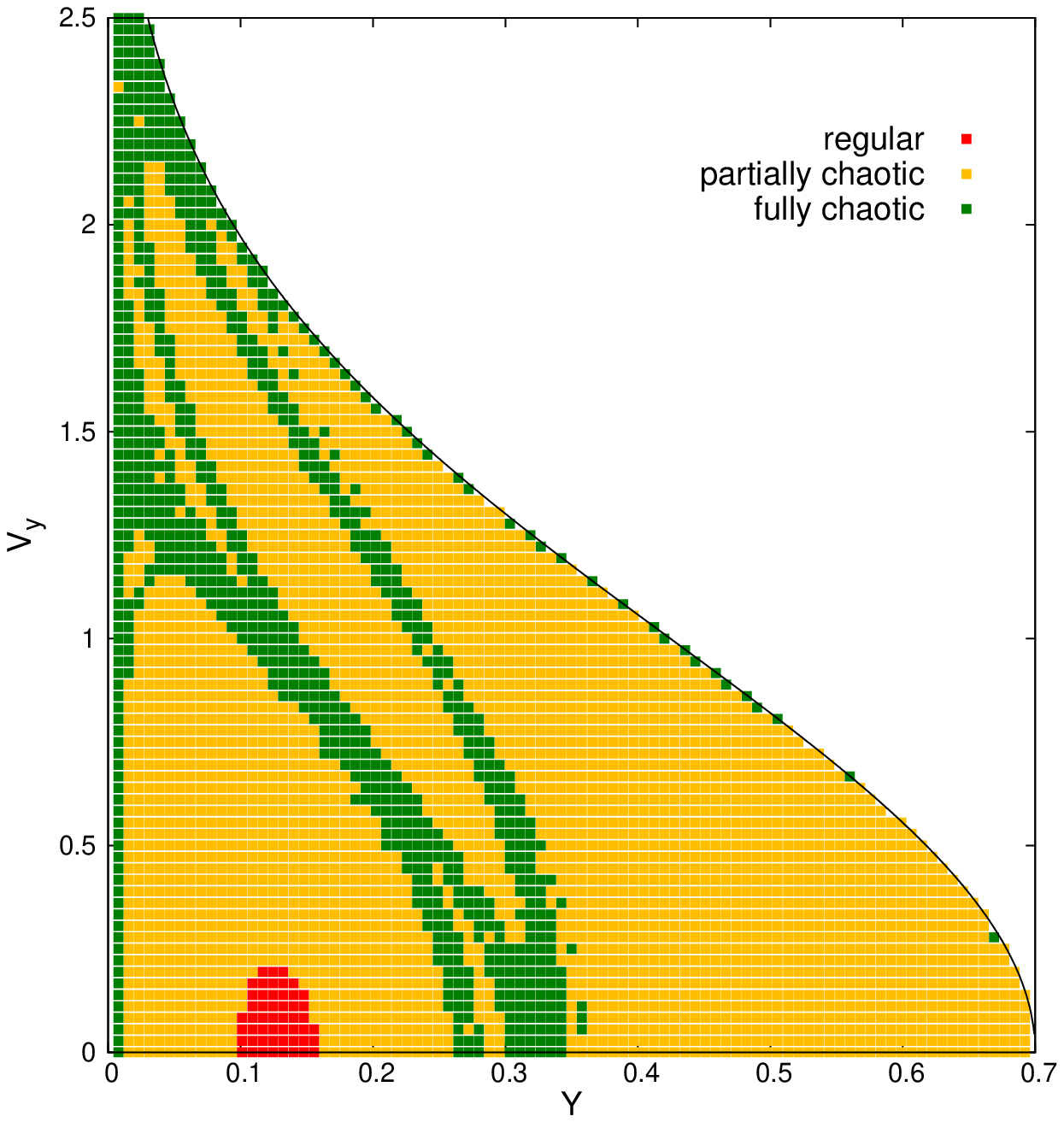}
 \caption{Left: Regular,
semi-chaotic or chaotic character of orbits in the 3D potential model with
$c=0.705$ when $z=0$. Right: the same, for the case $c=0.800$.}
 \label{fig:3D0705-08z0}
\end{figure}

We have already mentioned that, had we taken $z=0$ instead of $z=10^{-6}$ in the
initial conditions, the orbits could have not left the $(x, y)$ plane, but that
the chaotic character would be revealed anyway thanks to the variational
equations. The cases with $z=0$ for $c= 0.500, 0.695, 0.705$ and $0.800$ are
shown in Figs. \ref{fig:3D05-0695z0}, left and right, and \ref{fig:3D0705-08z0},
left and right, respectively. Comparison with the corresponding figures obtained
with an initial value $z=10^{-6}$ shows that the stochastic layers surrounding
the main resonances are always fully chaotic for those $c$ values. Those regions
only remained almost completely partially chaotic for low $c$ values (e.g.,
$c=0.05$ or $c=0.20$), for both initial $z$ values. When we reach $c=0.30$, the
plot for $z=0$ (not shown) does not differ much from that for $z=10^{-6}$ shown
in Fig. \ref{fig:2D-3D03} (right): There are only a few more partially chaotic,
rather than fully chaotic, orbits in the regions of the former minor
resonances. 

\section{Discussion}
\label{sec:discussion}

Our results confirm that, even for such a simple and well studied potential as
is the logarithmic one, 2D studies of the orbits in the principal symmetry
planes fail to reveal the chaotic nature of many orbits. A comparison of the 2D
results shown in Fig. \ref{fig:2D-3D03} (left) with the 3D results shown in all
the other figures reveals that most of the regular orbits in the 2D study become
either partially or fully chaotic in the 3D cases and that most of the 2D
partially chaotic orbits become fully chaotic in the 3D investigation. The
motions in the equatorial plane of very flat systems (e.g., $c=0.05$ or
$c=0.20$) seem to be the least affected ones by the third dimension, and the
reason is easy to understand.  The $x$, $y$ and $z$ accelerations are
proportional, respectively, to $x/a^{2}$, $y/b^{2}$, $z/c^{2}$ so that, all
other things been equal, smaller values of $c$ imply a much stronger force
normal to the principal plane and, therefore, a more stable motion in that
direction. It is interesting to note that, albeit for a different model,
\cite{FM97} also found that the stability of $(x, y)$ banana orbits improves for
lower $c/a$ values (see, e.g., their Figs. 7 and 8).

Regarding the nature of partially chaotic orbits, we notice that both
possibilities indicated in the Introduction can indeed take place. We notice
what clearly seems to be a chaotic layer made up of partially chaotic orbits
around the 1:1 loop resonance in Fig. \ref{fig:3D05-3D0695} (left) and around
the 2:1 banana resonance in Fig. \ref{fig:3D0705-3D08} (right); there is even a
mixture of partially and fully chaotic orbits around the 3:2 fish resonance in
Fig. \ref{fig:3D05-3D0695} (right). Alternatively, there is a connected area
dominated by partially chaotic orbits inside the 3:2 fish resonance in Fig.
\ref{fig:2D-3D03} (right), right in the middle of a region dominated by regular
orbits, and another region partially connected and partially scattered of
partially chaotic orbits in the middle of the fully chaotic domain in Fig.
\ref{fig:3D0705-3D08} (left).

Comparing the results obtained with the initial value $z=0$ with those with
$z(t=0)=10^{-6}$ we notice that essentially all the orbits that are regular in
one case are also regular in the other, so that the two isolating integrals that
they obey, besides energy, should be the same. For partially chaotic orbits,
instead, the story is different. A small fraction of those that appear in the
$z=0$ cases, remain as partially chaotic when $z(t=0)=10^{-6}$, but the bulk of
the partially chaotic orbits of the former case turn into fully chaotic orbits
in the latter one. Clearly, the isolating integral that sustains their partially
chaotic character when $z=0$ is destroyed when we let the orbit separate from
the principal plane of symmetry. Besides, such integral is not valid everywhere
because, except for cases with very low $c$ values, the connected region
in-between resonances is dominated by fully chaotic orbits, irrespectively of
whether the orbit remains in the main plane or not.

We may explore a little further the effect of the $z$ coordinate accepting that,
at least for small $z$ values, the $z$ motion should be oscillatory or near
oscillatory. Therefore, the additional isolating integral could be the
$z$-action or some related function. To check this, we computed for each orbit
the action
\begin{equation}
J_z={H\over\omega_z}={{1\over2}V_z^2+{V_0^2\over 2c^2}{z^2\over R_{\rm c}^2+
x^2/a^2 + y^2/b^2}\over {V_0\over c\sqrt{R_{\rm c}^2+x^2/a^2+y^2/b^2}}},
\end{equation}
where $\omega_z$ is the frequency of the $z$ motion computed by equating the
$z$-acceleration of the potential to that of an harmonic oscillator when $z\ll
1$, and $H$ is the harmonic oscillator Hamiltonian resulting from it. It turned
out that this quantity was not conserved for partially chaotic orbits, e.g., for
the case $c=0.3$, we found that all partially chaotic orbits had
$\log_{10}|J_{z\ {\rm final}}-J_{z\ {\rm initial}}|\in(-3,6)$ (we don't
normalize  because $J_z\simeq 0$).  Nevertheless, for the same $c=0.3$ case, all
regular orbits have $\log_{10}|J_{z\ {\rm final}}-J_{z\ {\rm
initial}}|\in(-12,-6)$, so that $J_z$, or some quantity related to it, is very
likely the isolating integral that distinguishes regular from partially chaotic
orbits in this case. The conservation of $J_z$ worsens as $c$ grows, however, so
that for larger $c$ values regular orbits should obey a different isolating
integral.

Therefore, our main conclusions are: 1) 3D studies are definitely necessary to
investigate the stability of (apparently) 2D orbits; 2) Instability normal to
the planes of symmetry worsens for lower values of the force normal to the plane
(i.e., larger $c$ values in the case of the logarithmic potential); 3) We find
partially chaotic orbits both in some stochastic layers around resonances and in
other regions (which may be coherent or not); 4) Although the isolating
integral, or pseudo integral, that distinguishes partially from fully chaotic
orbits in the cases investigated might depend on the separation from the main
plane ($z$), it is not related to the action normal to the plane; 5) The
isolating integral, or pseudo integral, that distinguishes regular from
partially chaotic orbits, instead, seems to be related to the action normal to
the plane, at least for low $c$ values.

\begin{acknowledgements}

We are very grateful to Ruben E. Mart\'{\i}nez and H\'ector R. Viturro for their
technical assistance. This work was supported with grants from the Consejo
Nacional de Investigaciones Cient\'{\i}ficas y T\'ecnicas de la Rep\'ublica
Argentina, the Agencia Nacional de Promoci\'on Cient\'{\i}fica y Tecnol\'ogica
and the Universidad Nacional de La Plata.

\end{acknowledgements}

\end{document}